\documentclass[review]{elsarticle}

\usepackage{graphicx}
\usepackage{epstopdf}
\usepackage{subcaption}
\usepackage{color}
\usepackage{rotating}

\journal{Journal of \LaTeX\ Templates}

\begin{document}

\begin{frontmatter}



\title{The effect of interplanetary magnetic field orientation on the solar wind flux impacting Mercury's surface}


\author[label1,label2,label3]{J. Varela}
\ead{deviriam@gmail.com (telf: 0033782822476)}

\author[label3]{F. Pantellini}
\author[label3]{M. Moncuquet}

\address[label1]{LIMSI, CNRS, Orsay, France}
\address[label2]{AIM DSM/IRFU/SAp, CEA Saclay, France}
\address[label3]{LESIA, Observatoire de Paris, CNRS, UPMC, Universite Paris-Diderot, 5
place Jules Janssen, 92195 Meudon, France}

\begin{abstract}

The aim of this paper is to study the plasma flows on the Mercury surface for different interplanetary magnetic field orientations on the day side of the planet. We use a single fluid MHD model in spherical coordinates to simulate the interaction of the solar wind with the Hermean magnetosphere for six solar wind realistic configurations with different magnetic field orientations: Mercury-Sun, Sun-Mercury, aligned with the magnetic axis of Mercury (Northward and Southward) and with the orbital plane perpendicular to the previous cases. In the Mercury-Sun (Sun-Mercury) simulation the Hermean magnetic field is weakened in the South-East (North-East) of the magnetosphere leading to an enhancement of the flows on the South (North) hemisphere. For a Northward (Southward) orientation there is an enhancement (weakening) of the Hermean magnetic field in the nose of the bow shock so the fluxes are reduced and drifted to the poles (enhanced and drifted to the equator). If the solar wind magnetic field is in the orbital plane the magnetosphere is tilted to the West (East) and weakened at the nose of the shock, so the flows are enhanced and drifted to the East (West) in the Northern hemisphere and to the West (East) in the Southern hemisphere.        

\end{abstract}


\end{frontmatter}


\section{Introduction}
\label{Introduction}

The Hermean magnetosphere is strongly dependent of the interplanetary magnetic field (IMF) module and orientation \cite{2007SSRv..132..529F}. The ratio between the intrinsic magnetic field of Mercury and the IMF is in the range $2$ to $10$ \cite{2011Sci...333.1859A}. The smallest values of the IMF module are below $10$ nT but reach $60$ nT when the planet is affect by a coronal mass ejection, with IMF orientations that differs from the Parker spiral \cite{2013JGRA..118...45B}. Other solar wind parameters as the electron density oscillates from $15$ to $160$ cm$^{-3}$, the temperature between $45,000$ to $160,000$ K, the velocity from $250$ to $600$ km/s and the $\beta = p/p_{mag}$ (ratio of the thermal to magnetic field strength) values between $0.08$ and $1$ \cite{2009JGRA..11410101B,2011PandSS...59.2066B}. The wide range of parameters leads to a large number of possible configurations of the solar wind (SW) and the Hermean magnetosphere \cite{2008Sci...321...82A,JGRE:JGRE3136}. In consequence, the plasma flows between the bow shock and the day side of the Hermean surface are very different depending on the SW conditions \cite{2012AGUFM.P32B..08B,2012LPI....43.1646D}. The stand off distance of the magnetopause and the IMF orientation are essential parameters to understand these flows. Previous studies pointed out the dependency of the particle precipitation on Mercury surface due to the IMF orientation and the stand off distance \cite{2003GeoRL..30.1877K,2003Icar..166..229M}. 

Several observational studies using MESSENGER data were performed by previous authors addressing the effect of the IMF orientation in the Hermean magnetosphere, analyzing the magnetic flux pile-up and the formation of a thick plasma depletion layer in the magnetosheat between the subsolar magnetopause and the bow shock \cite{2013JGRA..118.7181G}, the large magnetopause reconnection rate compared with the Earth for all IMF orientations due to the low beta of the solar wind in the inner heliosphere \cite{2009Sci...324..606S,2013AGUFMSM24A..03D} as well as the formation of a thick low-β plasma depletion layers in the inner magnetosheath adjacent to the subsolar magnetopause and a deeper magnetospheric cusp during a CME \cite{JGRA:JGRA51286}. The present research is devoted to complement previous simulations of the Hermean magnetosphere using single and multifluid codes \cite{2000Icar..143..397K,2008Icar..195....1K,2008JGRA..113.9223K}, hydrid simulations of Mercury’s dayside boundary layer \cite{Muller2011946,Muller2012666} and the interaction with the SW \cite{2012JGRA..11710228R,2010Icar..209...46W} as well as to compare with other terrestrial planet magnetospheres \cite{2008JGRA..113.6212J,Shimazu20081504}. We show how the different IMF orientations lead to an enhancement or a weakening of the Hermean magnetic field in distinct locations of the magnetosphere \cite{1979JGR....84.2076S}. In the magnetosphere regions where there is a strong reconnection between the IMF and the Hermean magnetic field the solar wind can be injected easily inside the inner magnetosphere, enhancing and shifting the local maximum of the fluxes towards the surface. The aim of this study is to characterize these flows on the day side of the planet for different IMF orientations.

We use a single fluid MHD model to simulate the interaction of the solar wind with the Hermean magnetic field. The code PLUTO \cite{2007ApJS..170..228M} is used in spherical coordinates for a realistic configuration of the solar wind obtained by the numerical models ENLIL + GONG WSA + Cone SWRC \cite{ODSTRCIL2003497,SWE:SWE449} and the IMF data from the MESSENGER magnetometer. For the Hermean magnetic field we use the model by Anderson et al 2012 \cite{2012JGRE..117.0L12A}. We perform six simulation with the IMF oriented in the Mercury-Sun (Bx case), Sun-Mercury (Bxneg), aligned with the magnetic axis in the Northward (Bz) and Southward (Bzneg) directions, and in the orbital plane of the planet perpendicular to the previous orientations (By and Byneg). 

The simulation results analysis is based on the next diagnostics:

- A comparison of the magnetic field module along the satellite trajectory with the MESSENGER data; a good agreement ensures the validity of the model and the results.

- Density distribution (polar cut); it shows the bow shock shape and the magnetospheric strucutres that link the back of the bow shock with the plane surface.

- Magnetic field distribution (frontal cut); indicates the reconnection regions and the magnetic field structure nearby the nose of the bow shock.

- Sinusoidal (Sanson-Flamsteed) projection of the inflow-outflow and open-close magnetic field lines regions in the planet surface; it shows the location of the regions with the strongest flows and the magnetic field configuration on the planet surface.

The paper is structured as follows. Section II, model description. Section III, simulations results. Section IV, conclusions.   

\section{Numerical model}
\label{Model}

We use the MHD version of the code PLUTO in spherical coordinates to simulate a single fluid polytropic plasma in the non resistive and inviscid limit. The code is freely available online \cite{2007ApJS..170..228M}.

The simulation domain is confined within two spherical shells centered in the planet, representing the inner and outer boundaries of the system. Between the inner shell and the planet surface (at radius unity in the domain) there is a "soft coupling region" where special conditions apply (defined in the next section).The shells are at $0.6 R_{M}$ and $12 R_{M}$ ($R_{M}$ is the Mercury radius).

The conservative form of the equations are integrated using a Harten, Lax, Van Leer approximate Riemann solver (hll) associated with a diffusive limiter (minmod). The divergence of the magnetic field is ensured by a mixed hyperbolic/parabolic divergence cleaning technique (DIV CLEANING) \cite{2002JCoPh.175..645D}.

The grid points are $196$ radial points, $48$ in the polar angle $\theta$ and $96$ in the azimuthal angle $\phi$ (the grid poles correspond to the magnetic poles).

The planetary magnetic field is axisymmetric with the magnetic potential $\Psi$ expanded in dipolar, quadrupolar, octupolar and 16-polar terms \cite{2012JGRE..117.0L12A}:

$$ \Psi (r,\theta) = R_{M}\sum^{4}_{l=1} (\frac{R_{M}}{r})^{l+1} g_{l0} P_{l}(cos\theta) $$  

The current free magnetic field is $B_{M} = -\nabla \Psi $. $r$ is the distance to the planet center and $\theta$ the polar angle. The Legendre polynomials in the magnetic potential $\Psi$ are:

$$ P_{1}(x) = x $$
$$ P_{2}(x) = \frac{1}{2} (3x^2 - 1) $$
$$ P_{3}(x) = \frac{1}{2} (5x^3 - 3x) $$
$$ P_{4}(x) = \frac{1}{2} (35x^4 - 30x^2 + 3) $$
the numerical coefficients $g_{l0}$ taken from Anderson et al. 2012 are summarized in the Table 1 \cite{2012JGRE..117.0L12A}.

\begin{table}[h]
\centering
\begin{tabular}{c | c c c c}
coeff & $g_{01}$(nT) & $g_{02}/g_{01}$ & $g_{03}/g_{01}$ & $g_{04}/g_{01}$  \\ \hline
 & $-182$ & $0.4096$ & $0.1265$ & $0.0301$ \\
\end{tabular}
\caption{Multipolar coefficients $g_{l0}$ for Mercury's internal field.}
\end{table}

The simulation frame is such that the z-axis is given by the planetary magnetic axis pointing to the magnetic North pole and the Sun is located in the XZ plane with $x_Sun > 0$. The y-axis completes the right-handed system.

\subsection{Boundary conditions and initial conditions}

The outer boundary is divided in two regions, the upstream part where the solar wind parameters are fixed and the downstream part where we consider the null derivative condition $\frac{\partial}{\partial r} = 0$ for all fields. At the inner boundary the value of the intrinsic magnetic field of Mercury is specified. In the soft coupling region the velocity is smoothly reduced to zero when approaching the inner boundary. The magnetic field and the velocity are parallel, and the density is adjusted to keep the Alfven velocity constant $v_{A} = B / \sqrt{\mu_{0}\rho} = 25$ km/s with $\rho = nm_{p}$ the mass density, $n$ the particle number, $m_{p}$ the proton mass and $\mu_{0}$ the vacuum magnetic permeability. In the initial conditions we define a paraboloid in the night side with the vertex at the center of the planet where the velocity is null and the density is two order smaller than in the solar wind. The IMF is cut off at $2 R_{M}$.

The solar wind parameters in the simulations are summarized in Table 2. We assume a fully ionized proton electron plasma, the sound speed is defined as $v_{s} = \sqrt {\gamma p/\rho} $ (with $p$ the total electron + proton pressure), the sonic Mach number as $M_{s} = v/v_{s}$ with $v$ the velocity and $M_{A} = v/v_{A}$ the Alfvenic Mach number. $\vec{v}_{u}$ is the unitary vector of the velocity.

\begin{table}[h]
\centering
\begin{tabular}{c | c c c c c}
Name & Date & B field (nT) & n (cm$^{-3}$) & $T$ (K) & $\beta$ \\ \hline
 Bx & 2012/01/19 & $(20,0,0)$ & $15$ & $85000$ & $0.11$\\
 Bxneg & 2011/10/17 & $(-18,0,0)$ & $20$ & $90000$ & $0.19$\\
 By & 2012/03/24 & $(-5,15,5)$ & $30$ & $100000$ & $0.38$\\
 Byneg & 2012/03/03 & $(8,-17,0)$ & $70$ & $115000$ & $0.79$\\
 Bz & 2011/09/06 & $(0,-10,41)$ & $90$ & $110000$ & $0.19$\\
 Bzneg & 2011/09/29 & $(8,10,-26)$ & $30$ & $60000$ & $0.07$\\
\end{tabular}
\caption{Simulations parameters I}
\end{table}

\begin{table}[h]
\centering
\begin{tabular}{c | c c c c}
Name & $v$ (km/s) & $\vec{v}_{u}$ & $M_{s}$  & $M_{A}$\\ \hline
 Bx & $320$ & $(-0.997,0.079,0)$ & $6.67$ & $2.83$\\
 Bxneg & $300$ & $(-0.994,0.112,0)$ & $6$ & $3.41$\\
 By & $400$ & $(-0.993,0.115,0)$ & $7.69$ & $6.06$\\
 Byneg & $380$ & $(-0.980,0.200,0)$ & $6.79$ & $7.75$\\
 Bz & $350$ & $(-0.986,0.165,0)$ & $6.4$ & $3.61$\\
 Bzneg & $360$ & $(-0.993,0.118,0)$ & $8.78$ & $3.13$\\
\end{tabular}
\caption{Simulations parameters II}
\end{table}

\section{Simulations results}
\label{Results}

The location of the bow shock and the magnetopause in the simulation for the IMF orientation Sun-Mercury and Mercury-Sun is similar in the simulation results and in MESSENGER data along the satellite trajectory, Fig. 1A (Bx) and 1B (Bxneg). There are closed field lines on the day side of the planet and magnetosphere structures linking the back of the bow shock with the planet surface near the poles, Fig. 1C (Bx) and 1E (Bxneg). There is a reconnection region in the South-East of the magnetosphere for the Bx IMF orientation and in the North-East in the Bxneg, Fig. 1D (Bx) and 1F (Bxneg), leading to an injection of solar wind plasma into the magnetosphere and down to the planetary surface \cite{1982GeoRL...9..921B}, Fig 1G (Bx) and 1H (Bxneg). The flux towards the planet surface for the Bx IMF case is enhanced in the South hemisphere compared with the Bxneg IMF hemisphere where the flux is enhanced in the North hemisphere. The open/closed magnetic field line regions (regions covered by blue dots in the figures) on the day and night side are wider in the Southern Hemisphere due to the Northward displacement of Mercury magnetic field, Fig. 1G and 1H. The plasma that enters in the inner magnetosphere through the reconnection regions follows the open magnetic field lines to the planet surface. The proportion of open/closed magnetic field lines regions on the night side are related with the magnetotail shape. In both simulations the magnetotail is slender and there are wide open magnetic field lines regions.

The location of the bow shock and the magnetopause along the satellite trajectory in the simulations with the IMF in the orbital plane is almost the same in the simulation results and the MESSENGER data, fig. 2A (By) and 2B (Byneg). The stand off distance of the bow shock is smaller compared with the Bx and Bxneg cases but there are still closed field lines on the day side of the planet, fig. 2C (By) and 2E (Byneg). This is consequence of the $\beta$ value, larger than in the previous simulations and the weakening of the planetary magnetic field due to the tilt of the magnetosphere, to the West in the By case (fig. 2D) and to the East in the Byneg case (fig. 2F). The reconnection regions in the equatorial plane of the planet are tilted in the same angle than the magnetosphere. The tilt of the magnetosphere leads to a East-West asymmetry in the distribution of the open magnetic field lines in the planet surface, fig. 2G (By) and 2H (Byneg), as well as the drift and enhancement of the fluxes local maximum on the surface correlated with the location of the reconnections. The open magnetic field line region on the night side are smaller due to the wide magnetotail. 

\begin{figure}[h]
\centering
\includegraphics[width=0.9\textwidth]{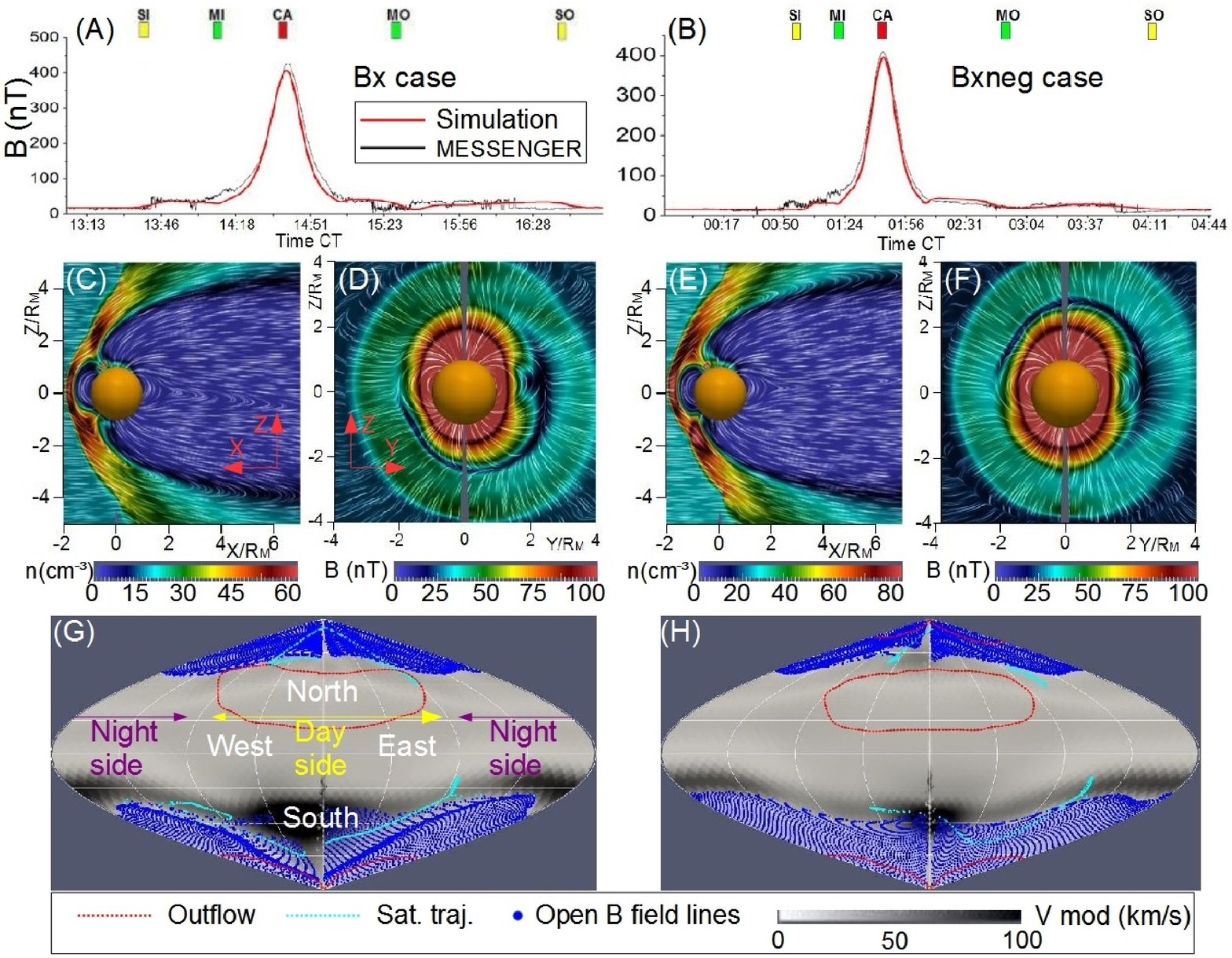}
\caption{Magnetic field along the satellite trajectory in the simulation compared with the MESSENGER data in the Bx (A) and the Bxneg (B) cases. Density distribution in a polar cut in the Bx (C) and Bxneg (E). Magnetic field distribution for a frontal cut in the Bx (D) and Bxneg (F) cases. We add the magnetic field lines in the distribution plots (white dashed lines). The inflow/outflow  and open/close magnetic field lines (blue dots) regions for the Bx (G) and Bxneg (H) cases. The red dotted lines encircle the regions with outflow. The light blue line shows the magnetic field lines connected with MESENGER trajectory. The polar cuts are displaced along the y axis by $0.1R_{M}$.}
\end{figure}  

There is a discrepancy in the bow shock location for the Bz simulation results and the MESSENGER data, fig. 3A. The mismatch is probably due to the fact that the satellite enters in the magnetosphere at the night side, nearby the outer boundary of the system where there is large effect of the boundary conditions in the bow shock shape. The MESSENGER data for the Bzneg orientation, fig. 3B, shows a rotation of the magnetic field components between the magnetopause and the satellite closest approach on the day side, observed in the magnetic field module as a local drop that is not reproduced in the simulation. 

\begin{figure}[h]
\centering
\includegraphics[width=0.9\textwidth]{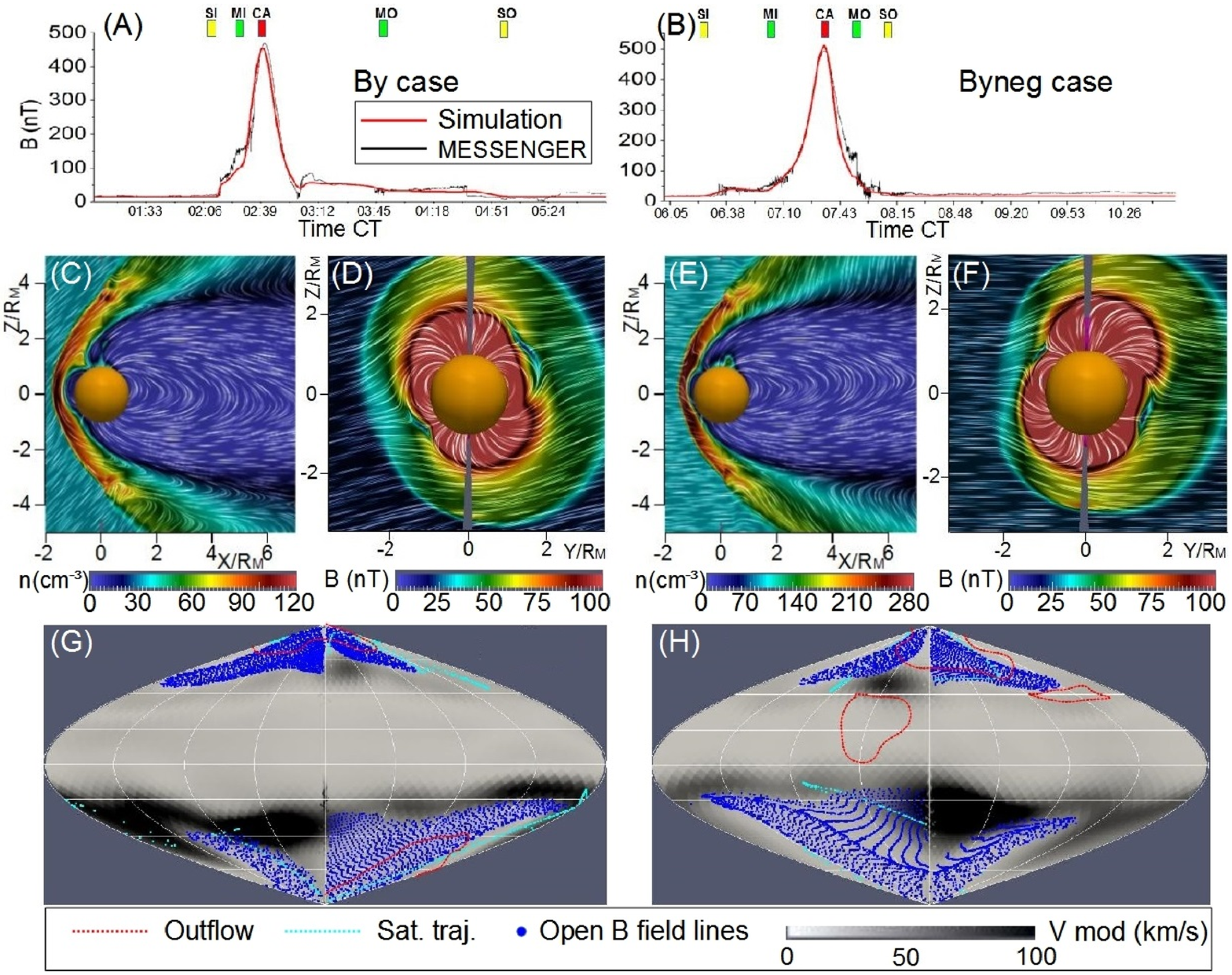}
\caption{Same format of Fig 1 for By and Byneg IMF.}
\end{figure} 

In both cases the location of the bow shock on the day side of the planet is almost the same comparing the simulation and the MESSSENGER data, SO in the Bz case and SI in the Bzneg simulation, suggesting that the magnetic structure and the flows on the planet surface should be similar. The simulation for the Bz IMF orientation, fig. 3C, shows close magnetic field lines in the day side but for the Bzneg case, fig. 3E, the back of the bow shock reaches the planet surface. The magnetic field is enhanced in the magnetosheath downstream of the bow shock for the Bz IMF orientation, fig. 3D, and weakened in the Bzneg simulation, fig. 3F.

\begin{figure}[h]
\centering
\includegraphics[width=0.9\textwidth]{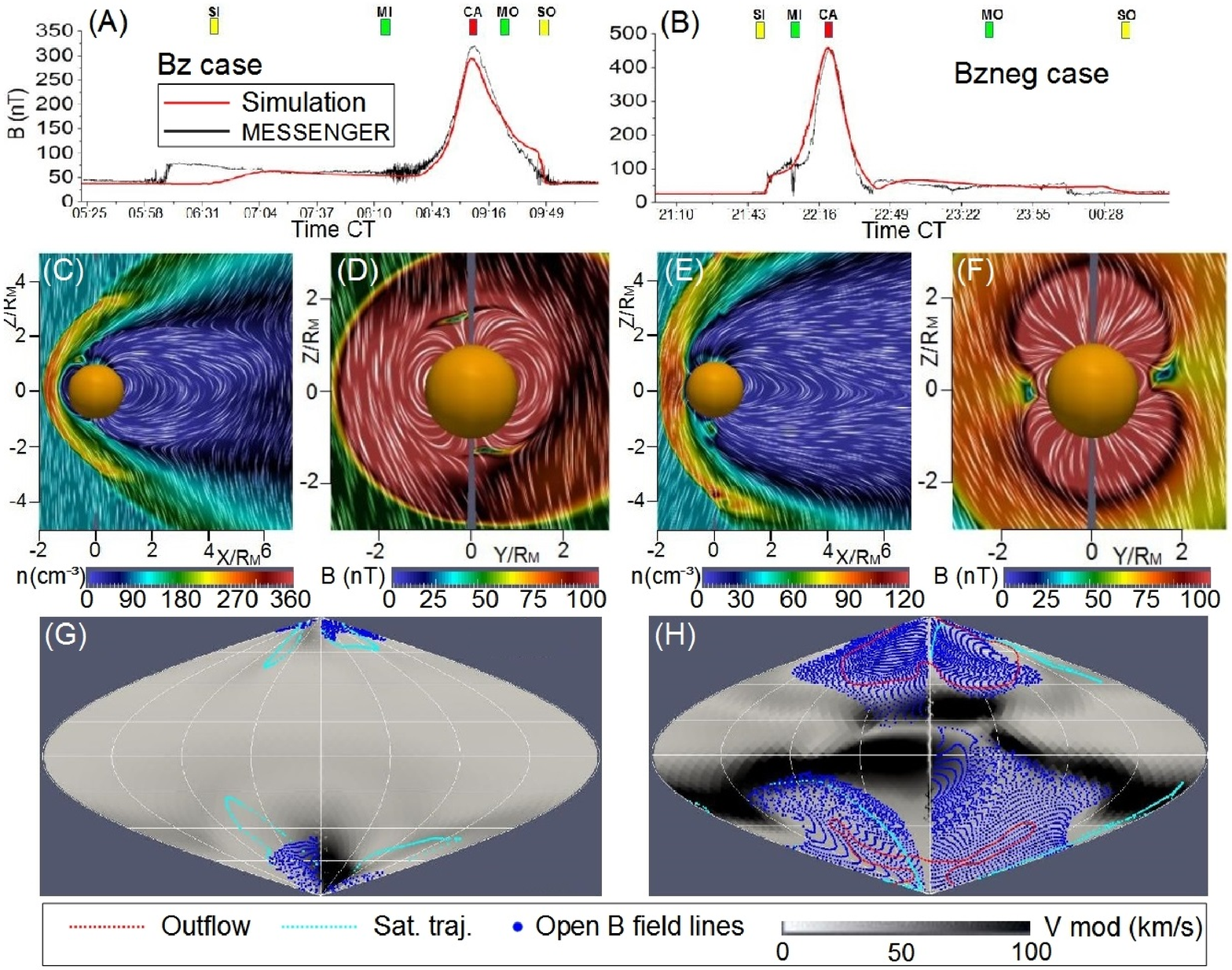}
\caption{Same format of Fig 1 for Bz and Bzneg IMF.}
\end{figure}  

There are reconnection regions nearby the North and South pole in the Bz case and at the equator for the Bzneg simulation. The regions with open magnetic field lines are small and located in the poles for the Bz case, fig. 3G, with the local maximum of the flow velocity enhanced and displaced to higher latitudes. The open magnetic field lines cover wide regions on the day side of the planet surface for the Bzneg simulation reaching low latitudes, enhancing the flows and drifting their maximum to the equator. The magnetotail is wide and the open magnetic field line regions are small on the night side of the planet. 

In summary the results show that the IMF orientation modifies the intensity of the downward flux of charged particles and where they impact the surface. The reason is the variation in the location of the reconnection sites on the magnetopause where magnetosheath plasma enters the magnetosphere. The configurations with a dominant Sun-Mercury (Mercury-Sun) IMF orientation show a reduction (enhancement) of the North/South asymmetry for the fluxes on the day side due to the Northward displacement of the planet magnetic field. The IMF orientations with a large component in the orbital plane perpendicular to the Sun-Mercury direction (By and Byneg) leads to a tilted magnetosphere with an East/West asymmetry that is observed too in the flow with their maximum displaced in the opposite sense. The Northward (Southward) IMF orientations show an increase (drop) of the planet magnetic field in the bow shock nose drifting to the poles and reducing (drifting to the equator and enhancing) the flows. The magnetotail shape is affected by the IMF orientation so the proportion of open/close magnetic field lines on the night side changes too in each simulation. If there are wide regions of open magnetic field lines as in the Mercury-Sun and Sun-Mercury orientations the Hermean surface on the night side is more exposed, but there are not plasma streams linking the back of the bow shock with the surface of the planet on the night side.

\section{Conclusions}
\label{Conclusions}

This study suggest that there is a direct correlation between the IMF orientation and the plasma flows from and towards the Mercury surface on the day side of the planet. A strong magnetic reconnection between the IMF and the planet magnetic field nearby the nose of the bow shock can enhance the flows and drifts their local maximum to different longitudes and latitudes. 

A IMF oriented in the Mercury-Sun direction leads to an enhancement of the North/South asymmetry for the flows on the planet surface due to the Northward displacement of the Hermean magnetic field. A wide reconnection region in the South of the magnetosphere nearby the nose of the bow shock enhances the flows on the South hemisphere and reduces them on the North Hemisphere. The opposite scenario is observed for the IMF orientation in the Sun-Mercury direction, with a wide reconnection region in the North of the magnetosphere that enhances the fluxes on the North Hemisphere while the flows in the South hemisphere are weakened.

For the Northward IMF orientation the planet magnetic field increases in the nose of the bow shock and there are small reconnection regions nearby the poles, leading to weaker fluxes and drifting their local maximum to the poles. The planet magnetic field for the Southward IMF orientation is weakened by the strong reconnection with the IMF and the back of the bow shock reaches the planet surface, leading to a large enhancement of the fluxes and the drift of the velocity local maximum to the equator.

If the IMF is mainly oriented in the orbital plane of the planet perpendicular to the Sun-Mercury direction, the magnetosphere is tilted and the planet magnetic field is weakened in the nose of the bow shock. The consequence is an East-West axymmetry and an enhanced of the fluxes through the reconnection regions nearby the bow shock nose that are tilted in the same angle than the magnetosphere.  

The IMF orientation affects the magnetotail shape and the proportion of open/close magnetic field lines on the night side of the planet. The planet surface on the night side is more exposed if the magnetotail is slender, with wilder regions of open magnetic field lines as can be observed in the Mercury-Sun and Sun-Mercury IMF orientations, as well as in the By and Byneg simulations due to the East/West asymmetry of the planet magnetic field.  

The simulations results are compatible with previous observational studies where a strong latitudinal variability in the surface flux is observed \cite{2012LPI....43.1646D}, the plasma precipitation in the region of the Hermean cusp to the planet surface and the depletion the of magnetosheath \cite{2003Icar..166..229M,2013JGRA..118.7181G}. The global magnetosphere structure in the present simulations is similar to the prediction of previous simulations \cite{2008Icar..195....1K,2012JGRA..11710228R}. Here we emphasize the role of the IMF orientation in the Hermean magnetosphere configuration, and how the location of the reconnection regions affects the plasma fluxes on the planet surface.

\section{Aknowledgments}
The research leading to these results has received funding from the European Commission's Seventh Framework Programme (FP7/2007-2013) under the grant agreement SHOCK (project number 284515). The MESSENGER magnetometer data set was obtained from the NASA Planetary Data System (PDS) and solar wind hydrodynamic parameters from the NASA Integrated Space Weather Analysis System.

Space Weather, 9, S03004, doi:10.1029/2011SW000663

\section*{References}

\bibliography{mybibfile}

\end{document}